\documentclass[twocolumn,twoside]{revtex4}
\usepackage{graphicx}
\usepackage{fancyhdr}
\def\ra{\rightarrow}

\def\babar{\mbox{{\normalsize \sl B}\hspace{-0.4em} {\small \sl A}\hspace{-0.03em}{\normalsize \sl B}\hspace{-0.4em} {\small \sl A\hspace{-0.02em}R}}}
\def\CP{$ C \! P$ }

\begin{document}

\title{Recent Results on $\tau$ Decays}

%

\author{G. Eigen\\
on behalf of the \babar\ Collaboration}
\affiliation{University of Bergen, Allegaten 55, 5007 Bergen, Norway}

\begin{abstract}
We present herein new results from Belle on the $\tau^- \ra \pi^- \nu_{\rm \tau} \ell^+ \ell^- $ branching fraction and from \babar\ on the $\tau^- \ra K^- (0,1,2,3) \pi^0 \nu_{\rm \tau}$,  $\tau^- \ra  \pi^- (3,4) \pi^0 \nu_{\rm \tau}$ and $\tau^- \ra K^- K^0_{\rm S} \nu_{\rm \tau}$ branching fractions. From the $K^- K^0_{\rm S}$
 mass spectrum we determine the spectral function. The improved branching fraction measurements of the $\tau^- \ra K^- (0,1,2,3) \pi^0 \nu_{\rm \tau}$ decays are used to determine  $|V_{\rm us}|$ from $\tau^- \ra X_{\rm s}^- \nu_{\rm \tau}$ inclusive decays.
\end{abstract}

\maketitle

\thispagestyle{fancy}


\section{Belle Measurement of the $\tau^- \ra \pi^- \nu_{\rm \tau} \ell^+ \ell^- $ Branching Fraction}

\subsection{Motivation}

\begin{figure}[h]
\centering
\includegraphics[width=75mm]{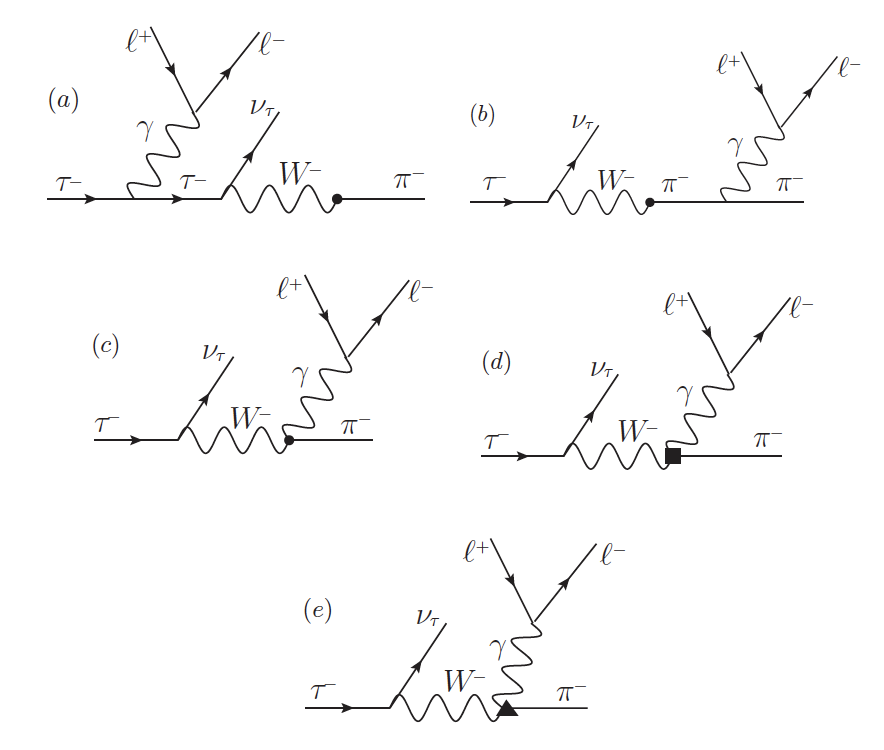}
\caption{Feynman diagrams for $\tau^- \ra \pi^- \nu_{\rm \tau} \ell^+ \ell^-$ decays, for (a) $\gamma$ emitted from the $\tau$, (b) $\gamma$ emitted from the $\pi$, (c) $\gamma$ emitted from $W \pi$ vertex, (d)  weak vector current coupling and (e) weak axial-vector current coupling.  }
\label{fig:qed}
\end{figure}

The decay $\tau^- \ra \pi^- \nu_{\rm \tau} \ell^+ \ell^-$~\cite{ccmode} with $\ell^+ \ell^- = e^+ e^-$ or $\mu^+ \mu^-$, receives QED contributions in which the photon is emitted from the $\tau$ and the $\pi$ shown in Figs.~\ref{fig:qed} (a, b, c) that are structure independent. In addition, weak contributions arise from the vector current and axial-vector current couplings shown in Figs.~\ref{fig:qed} (d, e) that are structure dependent. The last three diagrams involve a $\gamma^* W^* \pi$ vertex in which with two gauge bosons are off their mass shell. These couplings serve as a probe for new physics beyond the Standard Model (BSM). For example, a sterile $\nu$ that may explain MiniBooneÕs excess~\cite{miniboone} can enter the diagrams enhancing the branching fraction ${\cal B}(\tau^- \ra \pi^- \nu_{\rm \tau} \ell^+ \ell^-)$~\cite{dib}. If the photon is real, the  $\gamma W \pi$ vertex plays an important role for calculating radiative corrections for $\tau^- \ra \pi^- \nu_{\rm \tau}$, which helps with the evaluation of hadronic light-by-light scattering to the $(g-2)_{\rm \mu}$ calculation~\cite{guo, decker, miller}. Furthermore, ${\cal B}(\tau^- \ra \pi^- \nu_{\rm \tau} \ell^+ \ell^-)$ can be used to validate the Resonance Chiral Theory~\cite{ecker, cirigliano}. In the Standard Model (SM), the branching fraction predictions are ${\cal B}(\tau^- \ra \pi^- \nu_{\rm \tau} e^+ e^-)= (1.4-2.8) \times 10^{-5}$ and ${\cal B}(\tau^- \ra \pi^- \nu_{\rm \tau} \mu^+ \mu^-)= (0.03-1.0) \times 10^{-5}$~\cite{roig}.

\subsection{Analysis Strategy}

Belle performs a blind analysis using data recorded at the $\Upsilon(4S)$ with an integrated luminosity of $562~\rm fb^{-1}$. As a first step, they select $\tau^+ \tau^-$ events by requiring exactly four charged tracks with zero total charge. Each charged particle must have a transverse momentum greater than $p_{\rm T}> 0.1~\rm GeV/c$ and at least one charged particle must have $p_{\rm T}> 0.5~\rm GeV/c$. Isolated photons are required to have an energy of $E_{\rm \gamma} > 50~\rm ~ MeV ~(100~MeV)$ in the barrel (endcaps) to remove beam backgrounds. To reduce background contributions from radiative Bhabhas, $e^+ e^- \ra q \bar q$ ($q=u, d, s, c)$ and two-photon events, Belle requires the sum of the magnitude of momenta of the four charged particles to be in the range 3~ GeV/c $< \sum_{\rm i} |\vec p_{\rm i }| <$ 10 GeV/c, the missing mass to lie in the region 1 $\rm GeV/c^2 $ $< M_{\rm miss} <$ 7 $\rm GeV/c^2$ and a thrust of $0.85 < T < 0.99$.

The $\pi^- \nu_{\rm \tau} e^+ e^-$ signal sample is selected by identifying $e^+, e^-$ and $\pi^-$ using stringent particle identification criteria. To reduce hadronic background, the cosine of the angle between the $\tau^-$ and the $\pi^- e^+ e^-$ system is required to have $|\cos \theta_{\rm \tau, \pi e^+e^-} | < 1$. The main residual background comes from $ \tau^- \ra \pi^- \pi^0 \nu_{\rm \tau} e^+ e^-$ decays that have the same final state as the signal if a photon converts into $e^+ e^-$ or $\pi^0 $decays via the Dalitz mode. Thus, the $ee\gamma$ mass is required to lie outside the mass interval $110~\rm MeV/c^2 <$ $M_{\rm ee\gamma} < 165~\rm MeV/c^2$ and the transverse (longitudinal) decay length must be less than 1.2~cm (lie within the z-interval [-1.0~cm, 1.5~cm]).  The mass range $1.05~\rm GeV/c^2 <$$M_{\rm \pi ee} < 1.8~\rm GeV/c^2$ is chosen as signal region and $M_{\rm \pi ee} < 1~\rm GeV/c^2$ is defined as control region in which the 10243 observed events agree well with  $10083\pm 504$ expected background events. 

The $\pi^- \nu_{\rm \tau} \mu^+ \mu^-$ signal sample is selected by identifying both muons and the pion with stringent particle identification criteria, requiring high thrust and a di-muon mass of less than $\rm 0.85~GeV/c^2$ and selecting a pseudo tau mass 
\begin{eqnarray}
m^* = &\Big(&2 (E_{\rm \pi \mu \mu} - |\vec  p_{\rm \pi \mu\mu}|)(E_{\rm beam}-E_{\rm \pi \mu\mu})+M^2_{\rm \pi \mu \mu} \Big)^{0.5} \nonumber \\
 <&  &1.8~\rm  GeV/c^2
\label{eq:pseudomass}
\end{eqnarray}
where $E_{\rm beam}$ is the beam energy and $E_{\rm \pi \mu \mu}, \vec p_{\rm \pi \mu \mu}$ and $M_{\rm \pi \mu \mu}$ are energy, momentum and mass of the $\pi \mu^+ \mu^-$ system, respectively.
The main remaining background originates from $\tau^- \ra \pi^- \pi^+ \pi^- \nu_{\rm \tau}$ and $\tau^- \ra \pi^- \pi^+ \pi^- \pi^0 \nu_{\rm \tau}$  in which two oppositely-charged pions are  misidentified as  muons.  Since the muons come from pion decay in flight, many of the misidentified particles do not come from the interaction point and the transverse decay length provides a discriminating variable. The signal region is chosen as $R_{\rm xy} < 0.15 ~\rm cm$ and the control region as  $R_{\rm xy} > 0.20 ~\rm cm$. Belle observes 505 events in the control region that agrees with the $477\pm 22$ expected background events.

\begin{figure}[h]
\centering
\includegraphics[width=65mm]{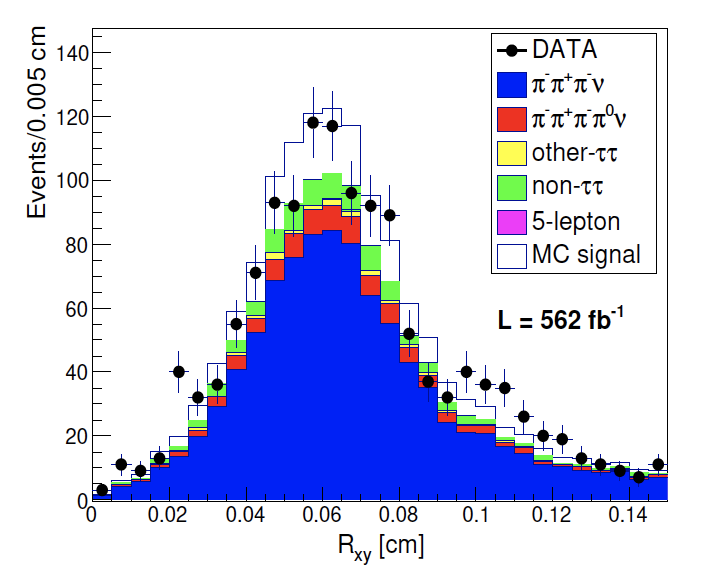}
\caption{The Belle preliminary $\pi ee$ invariant-mass distribution for $\tau^- \ra \pi^- \nu_{\rm \tau} e^+ e^-$ events.}
\label{fig:piee}
\end{figure}

\subsection{Results}

Figure~\ref{fig:piee} shows the  $\pi^- e^+e^-$ invariant mass for $\tau^- \ra \pi^- \nu_{\rm \tau} e^+ e^-$ events. In the signal region 676 events are observed compared to $478\pm 23$ events expected background. In the charge-conjugated mode the observed yield is 689 events compared to $476\pm 22$ expected background events. This provides a  $5.9\sigma$ significant excess.  The total systematic error is $14.4\%$ where the largest contribution arises from the particle identification efficiencies. With a signal efficiency of $\epsilon_{\rm sig}=1.88\pm 0.07\%$
Belle measures a preliminary branching fraction of ${\cal B}(\tau^- \ra \pi^- \nu_{\rm \tau} e^+ e^-) =(2.11\pm 0.19\pm 0.30) \times 10^{-5}$.

\begin{figure}[h]
\centering
\includegraphics[width=65mm]{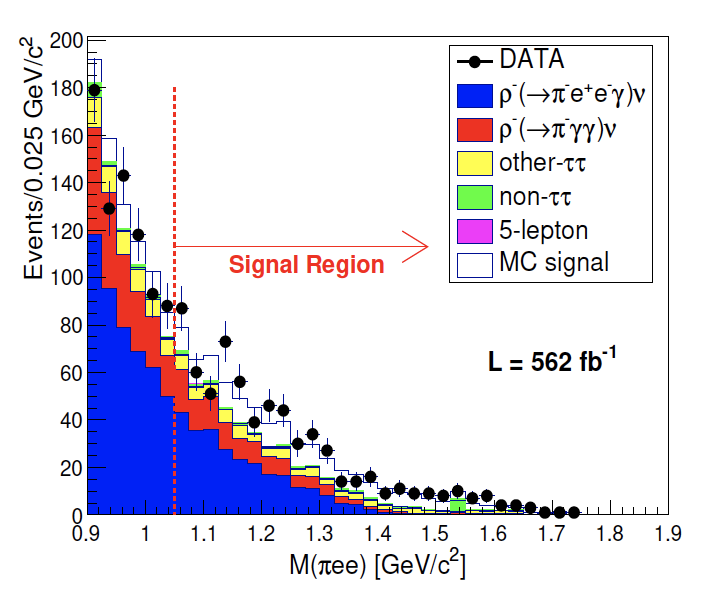}
\caption{The Belle preliminary transverse decay length distribution for $\tau^- \ra \pi^- \nu_{\rm \tau} \mu^+ \mu^-$ events.}
\label{fig:pimm}
\end{figure}

Figure~\ref{fig:pimm} shows the transverse decay length for $\tau^- \ra \pi^- \nu_{\rm \tau} \mu^+ \mu^-$ events in the signal region.  Belle observes 1315 events while the expected background is $1129 \pm 55$ events. In the charge-conjugated mode the yield is 1263 events compared to $1115\pm 54$ expected background events. The dominant backgrounds come from  $\tau^- \ra \pi^- \pi^+ \pi^- \nu_{\rm \tau}$ ($81.9\%)$ and $\tau^- \ra \pi^- \pi^+ \pi^- \pi^0 \nu_{\rm \tau}$  $(8.3\%)$. The total systematic uncertainty is $4.9\%$ where the largest contribution arises from particle identification. The detection efficiency is $4.14\pm0.16\%$. Since the total excess of $334\pm 51 \pm 109$ events has a statistical significance of $2.8\sigma$, Belle gives a preliminary branching fraction upper limit of ${\cal B}(\tau^- \ra \pi^- \nu_{\rm \tau} \mu^+ \mu^-) < 1.14  \times 10^{-5} ~@90\%$ confidence level (CL).

\section{\babar\ Study of $\tau^- \ra K^- (0-3) \pi^0 \nu_{\rm \tau}$  and $\tau^- \ra  \pi^- (3,4) \pi^0 \nu_{\rm \tau}$ Decays}

\subsection{Introduction}

The CKM element $|V_{\rm \rm us}|$ can be extracted from ${\cal B}(\tau \ra X_{\rm s} \nu_{\rm \tau})$ by~\cite{gamiz03, gamiz08}
\begin{equation}
|V_{\rm \rm us}| = \sqrt{ \frac{R_{\rm S}}{R_{\rm \rm V,A} |V_{\rm \rm ud}|^2 -\delta_{\rm theory}} }
\label{eq:vus}
\end{equation}
where
\begin{equation}
R_{\rm S}=\frac{{\cal B}(\tau \ra X_{\rm s}\nu_{\rm \tau})}{{\cal B}(\tau \ra e \bar \nu_{\rm e} \nu_{\rm \tau})},
\label{eq:rs}
\end{equation}
\begin{equation}
R_{\rm V,A}=\frac{{\cal B}(\tau \ra X_{\rm d}\nu_{\rm \tau})}{{\cal B}(\tau \ra e \bar \nu_{\rm e} \nu_{\rm \tau})},
\label{eq:rva}
\end{equation}
and $\delta_{\rm theory}$ represents the error from $SU(3)$ breaking effects. A  significant part of the experimental error on $|V_{\rm us}|$ results from the uncertainties on the $\tau^- \ra K^- (0-3)\pi^0 \nu_{\rm \tau}$ branching fractions.  In \babar\  we  measured the branching fractions of the decays $ \tau^- \ra K^-  (0-3) \pi^0 \nu_{\rm \tau}$ and $\tau^- \ra \pi^- (3,4) \pi^0 \nu_{\rm \tau}$. We use the decay modes $ \tau^-\ra \pi^- (0,1,2) \pi^0 \nu_{\rm \tau}$  and 
$\tau^- \ra \mu^- \bar \nu_{\rm \mu} \nu_{\rm \tau}$  as control samples. More precise branching fractions in these modes help to reduce the uncertainty on $|V_{\rm us}|$, since the inclusive branching fraction $\tau \ra X_{\rm s} \nu_{\rm \tau}$ is taken as a sum of exclusive $\tau$ decays with a kaon in the final state.

\begin{figure}[h!]
\centering
\includegraphics[width=70mm]{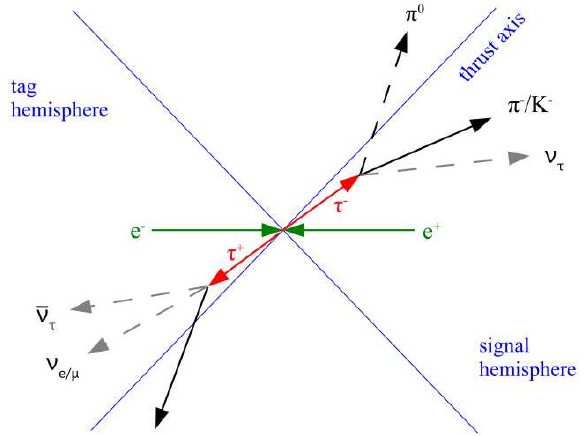}
\caption{  Event topology for $\tau^- \ra K^-,\pi^- \pi^0 \nu_{\rm \tau}$ with a lepton tag.}
\label{fig:tautau}
\end{figure}

\subsection{Analysis Method}

\begin{figure*}[]
\centering
\includegraphics[width=140mm]{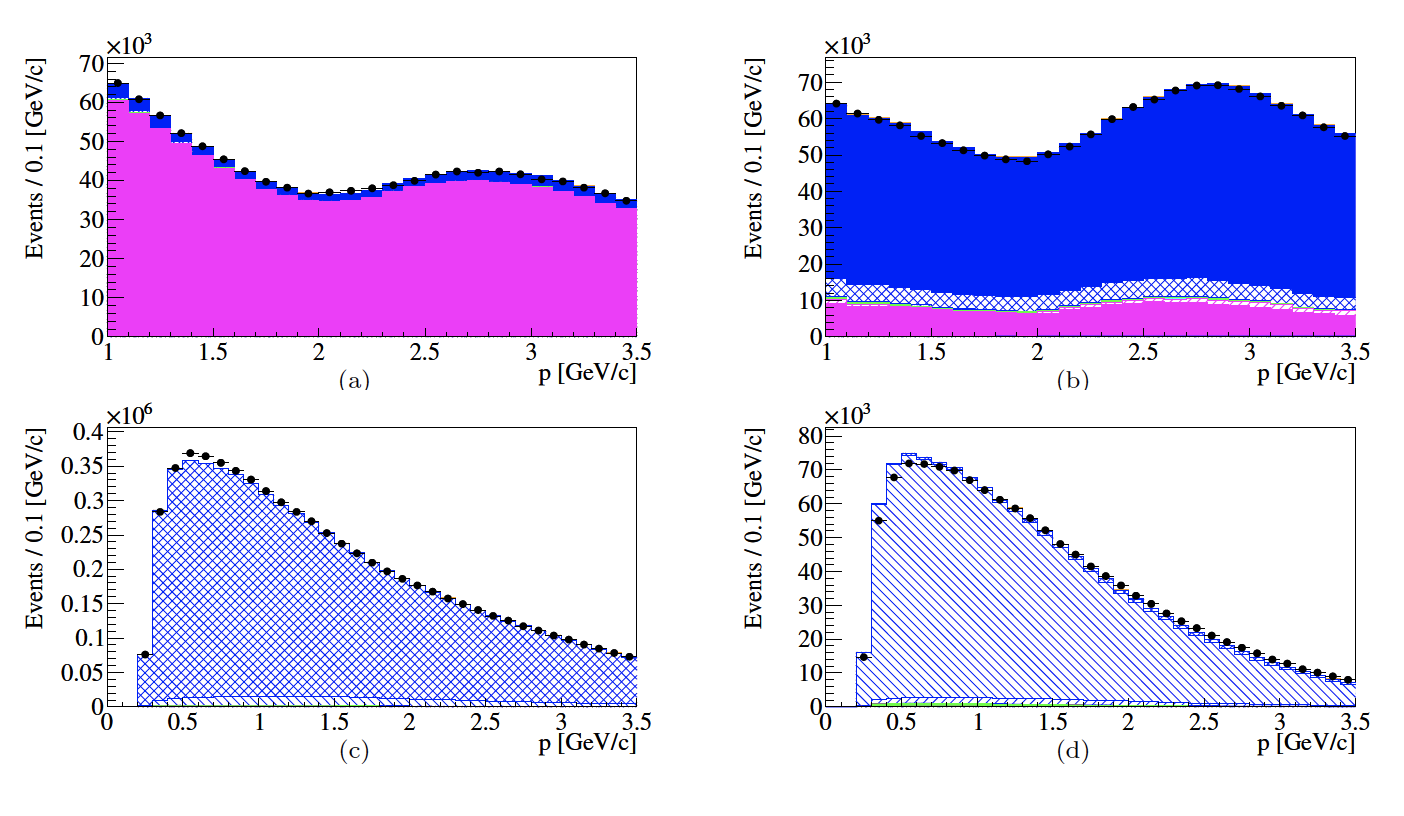}
\caption{\babar\ preliminary momentum  distributions of the charged particle in the signal hemisphere for selected candidates of the four control modes: (a) $\tau^- \ra \mu^- \bar \nu_{\rm \mu} \nu_{\rm \tau}$, (b)  $\tau^- \ra \pi^-  \nu_{\rm \tau}$, (c) $\tau^- \ra \pi^-  \pi^0 \nu_{\rm \tau}$ and (d) $\tau^- \ra \pi^-  2\pi^0 \nu_{\rm \tau}$. The individual contributions are shown in Fig.~\ref{fig:tauknp2}.}
\label{fig:tauknp1}
\end{figure*}
\begin{figure*}[]
\centering
\includegraphics[width=140mm]{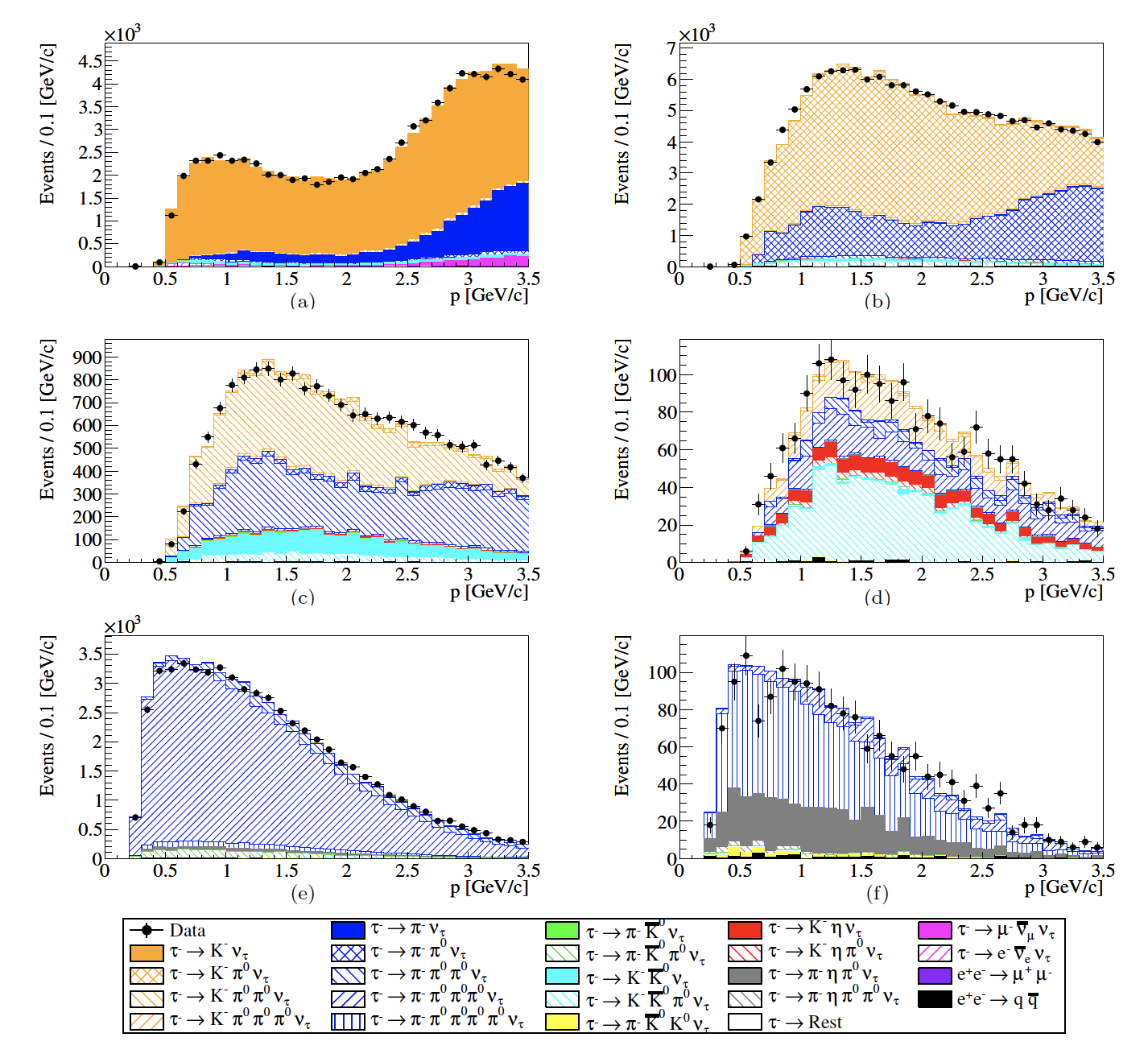}
\caption{\babar\ preliminary momentum  distributions of the charged particle in the signal hemisphere for the selected candidates of the six signal modes: (a) $\tau^- \ra K^-  \nu_{\rm \tau}$, (b)  $\tau^- \ra K^- \pi^0\nu_{\rm \tau}$, (c) $\tau^- \ra K^-  2\pi^0 \nu_{\rm \tau}$, (d) $\tau^- \ra K^-  3\pi^0 \nu_{\rm \tau}$ (e) $\tau^- \ra \pi^-  3\pi^0 \nu_{\rm \tau}$ and (f) $\tau^- \ra \pi^-  4\pi^0 \nu_{\rm \tau}$. }
\label{fig:tauknp2}
\end{figure*}

First, we divide the event into two hemispheres along the thrust axis as shown in Fig.~\ref{fig:tautau}. We tag $\tau^+$ in one hemisphere with $e^+$ or $\mu^+$ and select a $\pi^-$ or $K^-$ in the other hemisphere. We veto events that have additional charged particles. We keep all events that have $0-4~\pi^0$s with $\pi^0 \ra \gamma \gamma$  and veto events that have additional photons. To suppress two-photon events, we require the transverse momentum with respect to the missing energy to be
\begin{equation}
\frac{p_{\rm T}}{E_{\rm miss}}=\frac{\big( \vec p^{\rm ~CM}_{\rm sig} +\vec p^{\rm ~CM}_{\rm tag} \big)_{\rm T}}{\sqrt{s} -|\vec p^{\rm ~CM}_{\rm sig}|-|\vec p^{\rm ~CM}_{\rm tag}|} >0.2
\label{eq:pt}
\end{equation}
where $\sqrt{s}$ is the center-of-mass energy and  $p$ denotes the momenta of the signal and tag, respectively.
We suppress backgrounds with $K^0_{\rm L}$s by requiring that the missing-mass-squared 
\begin{equation}
m^2_{\rm miss} = p^2_{\rm miss} =\Big( p_{\rm e^+ e^-} - \sum_{\rm i} p_{\rm i}   \Big)^2  >0
\label{eq:mmiss}
\end{equation}
where $p_{\rm i}$ is the four-momentum of all reconstructed particles in the signal hemisphere. The explicit selection values are mode-specific.

We apply three corrections to the simulated data: a $\pi^0$ efficiency correction, a PID efficiency correction and a correction for neutron-induced showers..
For the $\pi^0$ efficiency correction we use the control samples and compare $\tau^- \ra \pi^- \nu_{\rm \tau}$ with  $\tau^- \ra \pi^- \pi^0 \nu_{\rm \tau}$ in data and Monte Carlo.
We define momentum-dependent correction factors, which are validated on the $ \tau^- \ra \pi^- 2 \pi^0 \nu_{\rm \tau}$ sample. For the PID efficiency correction
we use the $\tau^- \ra K^- K^+ \pi^- \nu_{\rm \tau}$ decay mode. We identify the $K^+$ and $\pi^-$ and test  particle identification on the $K^-$. 
Furthermore, we use the $\tau^- \ra \pi^-\pi^+\pi^- \nu_{\rm \tau}$ decay mode and identify both $\pi^-$ to test the PID on the $\pi^+$. We
measure the $K$ and $\pi$ PID efficiencies as functions of momentum, polar angle, azimuth angle, charge and \babar\ data-taking periods. 
Neutrons produced in hadron showers in the \babar\ electromagnetic calorimeter (EMC)  can  travel and produce a secondary shower that is
    identified as a photon. Since this process is not well-modeled in the MC, we have to apply a  correction.   For $\tau \ra \pi \nu_{\rm \tau}$, we see 
an enhancement  at small separations between a neutral shower and a charged pion in data, which is not seen in the MC. Thus, we define a weight to correct for this effect by comparing the number of events in data and MC for distances less than 40 cm yielding a correction of $w=0.972\pm 0.014$. 

\begin{figure*}[t!]
\centering
\includegraphics[width=140mm]{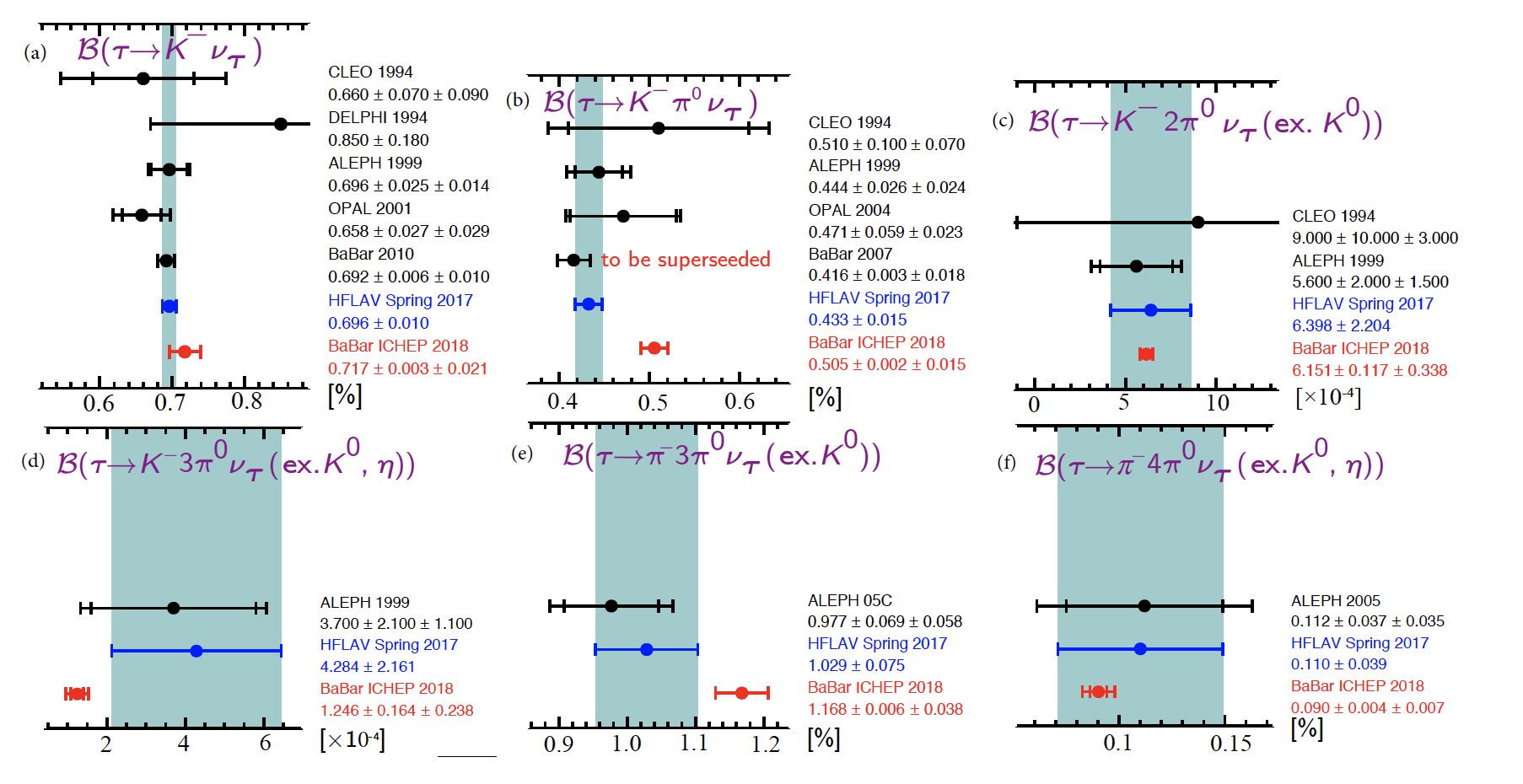}
\caption{\babar\ preliminary branching fractions for the six signal modes: (a) $\tau^- \ra K^-  \nu_{\rm \tau}$, (b)  $\tau^- \ra K^- \pi^0 \nu_{\rm \tau}$, (c) $\tau^- \ra K^-  2\pi^0 \nu_{\rm \tau}$, (d) $\tau^- \ra K^-  3\pi^0 \nu_{\rm \tau}$ (e) $\tau^- \ra \pi^-  3\pi^0 \nu_{\rm \tau}$ and (f) $\tau^- \ra \pi^-  4\pi^0 \nu_{\rm \tau}$. }
\label{fig:tauknp9}
\end{figure*}

\subsection{Results}

Figures~\ref{fig:tauknp1} show the momentum distributions of the charged particle in the signal hemisphere for the selected candidates of the four control modes. The data are well described by the simulation. Figures~\ref{fig:tauknp2} show the corresponding momentum distributions  of the six signal modes. Again, data are well described by simulations.

To determine the branching fractions, we need to account for cross feeds among the six signal modes. Using simulation, we first subtract in each observed channel $N_{\rm j}^{\rm obs}$ all backgrounds $N_{\rm j} ^{\rm bkg}$  that do not originate from the six signal channels. Then, we determine the migration matrix $M_{\rm ij}$, which gives the probability that a produced mode $i$ is observed in mode $j$.
Inversion of the matrix yields the number of truly produced events
\begin{equation}
N_i^{\rm prod} = \Big( M^{ -1}   \Big)_{\rm ij} \Big( N^{\rm obs}_{\rm j} -N_{\rm j}^{\rm bkg} \Big).
\label{eq:nprod}
\end{equation}    
The branching fractions are calculated by
\begin{equation}
{\cal B}(\tau \ra i) = 1 - \sqrt{1-\frac{N^{\rm prod}_i}{{\cal L} \sigma_{\rm \tau \tau}} }
\label{eq:bknp}
\end{equation}  
where ${\cal L}= 473.9~\rm fb^{-1}$ is the integrated luminosity near or at the $\Upsilon(4S)$ and $\sigma_{\rm \tau \tau}=0.919\pm 0.003~\rm nb$ is the $e^+ e^- \ra \tau^+ \tau^-$ cross section at 10.58~GeV. For the $\tau^- \ra K^- (\pi^-) n\pi^0 \nu_{\rm \tau}$ modes the efficiencies are in the 0.1 to 2\% (0.1 to 3.3\%) range while the efficiency for $\tau^- \ra \mu^- \bar \nu_{\rm \mu} \nu_{\rm \tau}$ is 1.3\%. 

For the six signal modes we measure the following branching fractions
\begin{eqnarray}
&{\cal B} (\tau^- \ra K^- \nu_{\rm \tau}) &=(7.17 \pm 0.03 \pm 0.21) \times 10^{-3}  \nonumber \\
&{\cal B} (\tau^- \ra K^- \pi^0 \nu_{\rm \tau}) &=(5.05 \pm 0.02\pm 0.15 ) \times 10^{-3}\nonumber \\
&{\cal B} (\tau^- \ra K^- 2\pi^0 \nu_{\rm \tau}) &=(6.15 \pm 0.12\pm 0.33)\times 10^{-4} \nonumber \\
&{\cal B} (\tau^- \ra K^- 3\pi^0 \nu_{\rm \tau}) &=(1.25 \pm 0.16\pm 0.24)\times 10^{-4} \nonumber \\
&{\cal B} (\tau^- \ra \pi^- 3\pi^0 \nu_{\rm \tau}) &=(1.17\pm 0.01\pm 0.04)\times 10^{-2} \nonumber \\
&{\cal B} (\tau^- \ra \pi^- 4\pi^0 \nu_{\rm \tau}) &=(9.02 \pm 0.40\pm 0.65)\times 10^{-4}.~~~~
\label{eq:bfknp}
\end{eqnarray}
The first error is statistical and second systematic.
Figure~\ref{fig:tauknp9} shows the branching fractions measured by \babar\ together with previous results.
The branching fraction for $\tau^- \ra K^- \nu_{\rm \tau}$  is slightly worse than that obtained with a three-prong tag~\cite{babar10}, while the branching fraction for 
 $\tau^- \ra K^- \pi^0 \nu_{\rm \tau}$ is much improved. The branching fractions for the other four modes are the first \babar\ measurements. They are also much more precise than previous results.

\section{\babar\ Measurement of the branching fraction and Spectral Function of $\tau^- \ra K^- K^0_{\rm S} \nu_{\rm \tau}$ }

\subsection{Motivation}

\babar\ used the decay $\tau^- \ra K^- K^0_{\rm S} \nu_{\rm \tau}$ to measure the 
spectral function in this channel~\cite{tsai71}
\begin{equation}
V(q)=\frac{m_{\rm \tau}^8}{12 \pi C(q) |V_{\rm ud}|^2} 
\frac{{\cal B}(\tau^- \ra K^- K^0_{\rm S} \nu_{\rm \tau})}
{{\cal B}(\tau^- \ra e^- \bar \nu_{\rm e} \nu_{\rm \tau})}\frac{1}{N} \frac{dN}{dq}
\label{eq:specfun}
\end{equation}  
where $m_{\rm \tau}$ is the $\tau$ mass, $q$ is the invariant mass of the $K^- K^0_{\rm S}$ system, $V_{\rm ud}$ is a CKM matrix element, $(dN/dq)/N$ is the normalized 
 $K^- K^0_{\rm S}$ mass spectrum and $C(q)$ is a phase space factor
\begin{equation}
C(q)=q(m^2_{\rm \tau} -q^2)^2(m^2_{\rm \tau} +2 q^2).
\label{eq:phasespace}
\end{equation}   
Since  the vector current is conserved~\cite{tsai71}, the same spectral function appears in the isovector part of the $e^+ e^- \ra K \bar K$ cross section
\begin{equation}
\sigma^{\rm I=1}_{\rm e^+ e^- \ra K\bar K} (q)=\frac{4 \pi^2 \alpha^2}{q^2} V(q),
\label{eq:specfun}
\end{equation}  
where $\alpha$ is the fine structure constant. \babar\ measured the cross sections for $e^+ e^- \ra K^+ K^-$ and  $e^+ e^- \ra K^0_{\rm S} K^0_{\rm  L}$~\cite{babar13a, babar13b}. In addition, SND measured the cross section for  $e^+ e^- \ra K^+ K^-$~\cite{snd16}.
Combining the data of both experiments we can determine the  moduli of isovector and isoscalar form factors and relative phase between them in a model-independent way. While Belle measured the branching fraction for $\tau^- \ra K^- K^0_{\rm S} \nu_{\rm \tau}$ rather precisely ($3\%$)~\cite{belle14}, CLEO measured the $K^- K^0_{\rm S}$ mass spectrum~\cite{cleo96} with large uncertainties.


\subsection{Analysis strategy}

Using an integrated luminosity of ${\cal L} = (468\pm 2.5)~\rm  fb^{-1}$ \babar\ has studied $ \tau^- \ra K^- K^0_{\rm S} \nu_{\rm \tau}$. As in the analysis above, we divide  the event into two hemispheres. On the tag side we require an identified electron or muon. The center-of-mass momentum of the lepton tag must lie between $\rm 1.2~GeV/c$ and $\rm 4.5~GeV/c$ with a polar angle satisfying $|\cos \theta_{\rm \ell}| < 0.9$. This removes QED events $e^+ e^- \ra e^+ e^-, \mu^+ \mu^-$. On the signal side  we select 
$ \tau^- \ra K^- K^0_{\rm S} \nu_{\rm \tau}$ by requiring an identified $K^-$ and two oppositely-charged pions that are compatible with a $K^0_{\rm S}$ decay, having a
decay length larger than $1~\rm cm$ and a mass consistent with the nominal $K^0_{\rm S}$ mass. To suppress background from charged  pions we require the charged $K$ momentum to satisfy $\rm 0.4~GeV/c <$ $p_{\rm K} < $ $\rm  5.0~GeV/c$ and the polar angle to satisfy $-0.7374 < \cos \theta_{\rm K} < 0.9005$.
In addition to other standard selection criteria, the sum of photon energies has to be less than 2~GeV~\cite{babar12, babar18}. The selection reduces the $\tau $ and $ q \bar q)$ backgrounds by 3.5 and 5.5 orders of magnitude, respectively. We determine the non-$K^0_{\rm S}$ background from the $m_{\rm \pi^+ \pi^-}$ sidebands and perform a bin-by-bin subtraction in the $m_{\rm K^- K^0_{\rm S}}$ mass spectrum. The background fraction is of order $10\%$ for $m_{\rm K^- K^0_{\rm S}} < 1.3~\rm GeV/c^2$ increasing to $50\%$ for masses above $\rm 1.6~GeV/c^2$. The $\tau^+ \tau^-$ background consists of $\sim 80\%$  $\tau^- \ra K^- K^0_{\rm S} \pi^0 \nu_{\rm \tau}$, $10\% $ $\tau^- \ra \pi^- K^0_{\rm S}  \nu_{\rm \tau}$ and $3\%$ $\tau^- \ra \pi^- K^0_{\rm S}\pi^0 \pi^0 \nu_{\rm \tau}$. The remaining background comes from a misidentified lepton on the tag side. For subtraction of background without $\pi^0$s we use simulation. For background with $\pi^0$s, we perform a bin-by-bin subtraction. We divide the data into two classes, one without  $\pi^0$ and one with one $\pi^0$s. 
\begin{eqnarray}
N_{\rm 0\pi^0} &=&(1-\epsilon_{\rm s})N_{\rm s} +(1-\epsilon_{\rm b})N_{\rm b} \nonumber \\
N_{\rm 1\pi^0}&=& \epsilon_{\rm s} N_{\rm s} + \epsilon_{\rm b} N_{\rm b} ~~~
\label{eq:bkg}
\end{eqnarray}  
where $N_{\rm 0\pi^0}$ and $N_{\rm 1\pi^0}$ are the number of selected data events without a $\pi^0$ and with a $\pi^0$ and $\epsilon_{\rm s}$ and $\epsilon_{\rm b}$ are the probabilities for signal and background $\tau^+ \tau^-$ events to be observed in the class with one $\pi^0$. The probabilities are determined from MC as a function of $m_{\rm K^- K^0_{\rm S}}$ bins. We then correct the value of $\epsilon_{\rm b}$ by the $\pi^0$ efficiency correction of $0.984\pm 0.006$.  We need to adjust $\epsilon_{\rm s}$ by $1.05\pm 0.05$. With the corrected values we determine $N_{\rm s}$ and $N_{\rm b}$. The selection efficiency as a function of $m_{\rm K^- K^0_{\rm S}}$ is about $13\%$ at low masses decreasing to $11\%$ at high masses. The total systematic uncertainty is $2.7\%$ where the largest contribution comes from the background with one $\pi^0$.

\subsection{Results}

\begin{figure}[h]
\centering
\includegraphics[width=70mm]{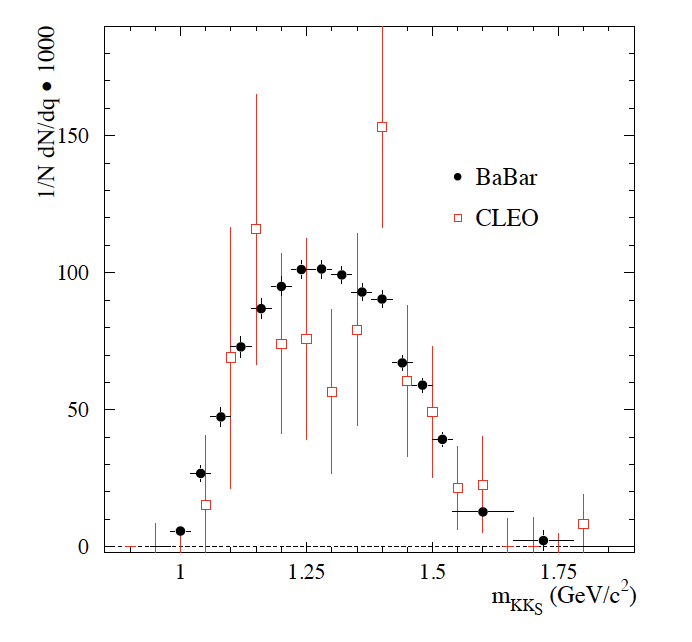}
\caption{Normalized $K^- K^0_{\rm S}$ invariant-mass spectrum of the $\tau^- \ra K^- K^0_{\rm S} \nu_{\rm \tau} $ decay for \babar\ (solid points with error bars) and for CLEO (open squares with error bars). The errors are only statistical.
 }
\label{fig:mtkk}
\end{figure}

\begin{figure}[h]
\centering
\includegraphics[width=70mm]{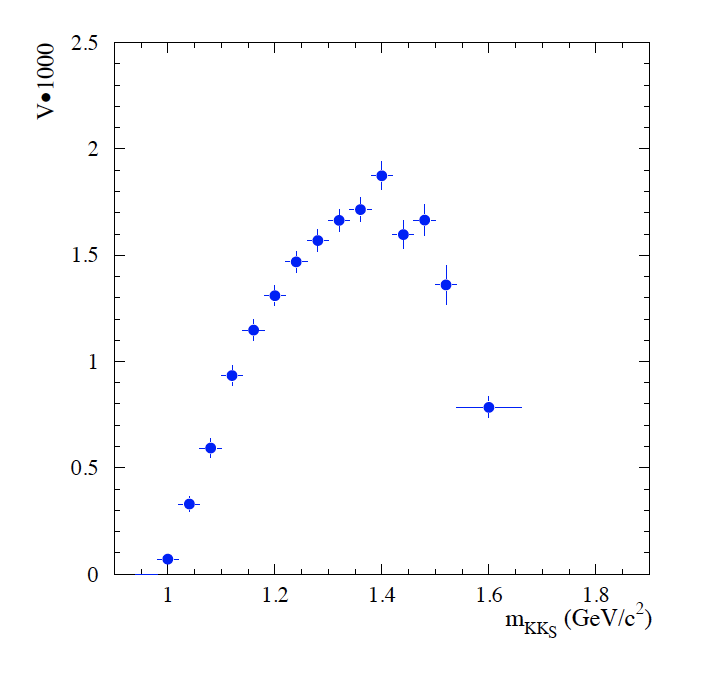}
\caption{Measured spectral function for $\tau^- \ra K^- K^0_{\rm S} \nu_{\rm \tau} $. Errors are statistical only. 
 }
\label{fig:specfun}
\end{figure}

The $\tau^- \ra K^- K^0_{\rm S} \nu_{\rm \tau}$ branching fraction  is obtained from
\begin{equation}
{\cal B}(\tau^- \ra K^- K^0_{\rm S} \nu_{\rm \tau}) =\frac{N_{\rm exp}}{2 {\cal L} B_{\rm lep} \sigma_{\rm \tau \tau}}
\label{eq:btkk}
\end{equation}  
where $B_{\rm lep}=0.3521 \pm 0.0006$ is the world average of the combined $\tau \ra e \bar \nu_e \nu_{\rm \tau}$ and $\tau \ra \mu \bar \nu_\mu \nu_{\rm \tau}$ branching fractions~\cite{PDG}. We observe a total number of $N_{\rm s}=223741\pm 3461$ signal events yielding
\begin{equation}
{\cal B}(\tau^- \ra K^- K^0_{\rm S} \nu_{\rm \tau}) =(0.739\pm 0.011\pm 0.020)\times 10^{-3}
\label{eq:btkk2}
\end{equation}  
Our result agrees well with the Belle measurement of $(0.740\pm 0.007\pm 0.027) \times 10^{-3}$~\cite{belle14}. Figure~\ref{fig:mtkk} shows the normalized $K^- K^0_{\rm S}$ invariant-mass spectrum for $\tau^- \ra K^- K^0_{\rm S} \nu_{\rm \tau}$ from which the spectral function shown in Fig.~\ref{fig:specfun} is extracted.  Our $K^- K^0_{\rm S}$ invariant-mass spectrum is much more precise than the one from CLEO~\cite{cleo96}. 


\section{Measurement of $|V_{\rm us}|$ in inclusive $\tau^- \ra X^-_{\rm S} \nu_{\rm \tau}$ Decays}

\begin{figure}[t!] 
\centering
\includegraphics[width=70mm]{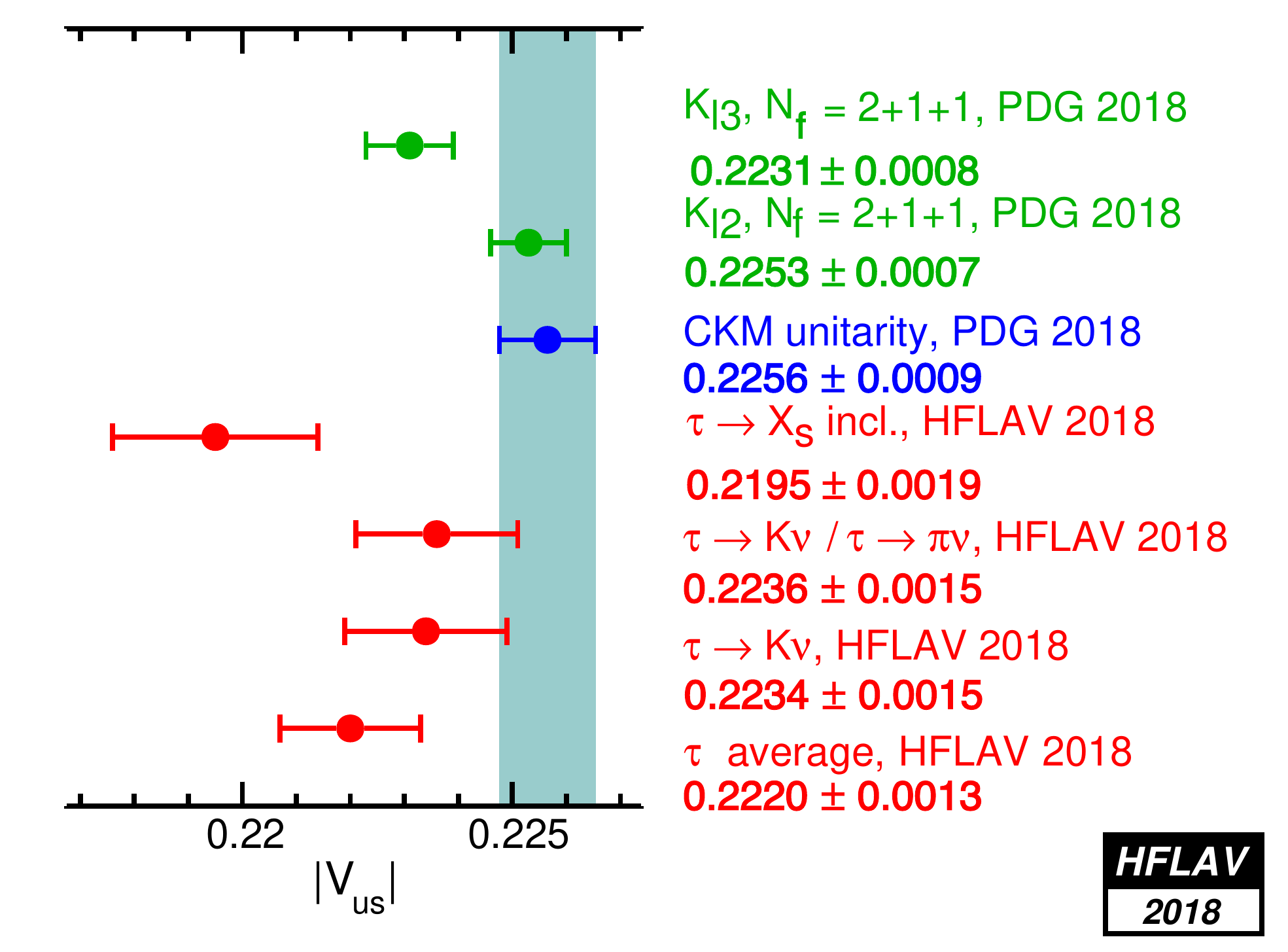}
\caption{ Compilation of $|V_{\rm us}|$ measurements from $K_{\rm \ell 3}$ and $K_{\rm \ell 2}$ decays (green points)~\cite{antonelli10, PDG}, CKM unitarity (blue point)~\cite{PDG}, new $\tau^- \ra X^-_{\rm S} \nu_{\rm \tau}$ inclusive analysis (upper red point)~\cite{gamiz03, gamiz08}, two determinations of $\tau^- \ra K^- \nu_{\rm \tau}$ (two middle red points)~\cite{HFLAV} and $\tau$ average (lower red point)~\cite{HFLAV}.
 }
\label{fig:vus}
\end{figure}

The CKM element $|V_{\rm us}|$ is typically determined from $K_{\rm \ell 2}$ and $K_{\rm \ell 3}$ decays. Using the unitarity of the CKM matrix  $|V_{\rm us}|$ is determined with the smallest uncertainty since $|V_{\rm ud}|$ is the best measured CKM element with an uncertainty of 0.02\% and the contribution of $|V_{\rm ub}|$ is negligible despite its large uncertainty. Furthermore, we can determine $|V_{\rm us}|$ from $\tau$ decays with a kaon in the final state.
Figure~\ref{fig:vus} shows  $|V_{\rm us}|$ extracted from inclusive $\tau^- \ra X_{\rm s}^- \nu_{\rm \tau}$ decays in comparison with results from $\tau^- \ra K^- \nu_{\rm \tau}$, $K_{\rm \ell 3 }$ and $K_{\rm \ell 2}$ decays, and CKM unitarity~\cite{HFLAV, ckm, PDG}. The $|V_{\rm us}|$ value from  inclusive $\tau^- \ra X_{\rm s}^- \nu_{\rm \tau}$ decays lies $-2.9\sigma$ lower than the result from CKM unitarity. With the new $\tau^- \ra K^- (0-3) \pi^0 \nu_{\rm \tau}$ \babar\ branching fraction measurements the precision on $|V_{\rm us}|$ improved though the discrepancy changed only slightly from the previous HFLAV analysis yielding $-3.0\sigma$. The value of $|V_{\rm us}|$ extracted from the previous \babar\ $\tau^- \ra K^- \nu_{\rm \tau}$ measurement is consistent with  $|V_{\rm us}|$ determined from CKM unitarity within $2\sigma$.  In the inclusive $\tau^- \ra X^-_{\rm s} \nu_{\rm \tau}$ analysis the precision can be further improved by remeasuring other decay modes more precisely that enter the inclusive $\tau^- \ra X^-_{\rm s} \nu_{\rm \tau}$ analysis, such as $\tau^- \ra \pi^- \bar K^0_{\rm S} 2\pi^0 \nu_{\rm \tau}$, $\tau^- \ra \bar  K^0_{\rm S} \pi^- \pi^+ \pi^- \nu_{\rm \tau}$, $\tau^- \ra \bar  K^- \pi^- \pi^+ \pi^0 \nu_{\rm \tau}~(ex. ~K^0_{\rm S}, \omega, \eta)$, $\tau^- \ra \bar  K^0_{\rm S}\pi^-\nu_{\rm \tau}$, $\tau^- \ra \bar  K^0_{\rm S} \pi^- \pi^0\nu_{\rm \tau}$ and $\tau^- \ra \bar  K^- \omega \nu_{\rm \tau}$. Other approaches are based on using precise kaon decay branching fractions to predict $\tau $ decay branching fractions or use the $\tau$ spectral functions to extract $|V_{\rm us}|$~\cite{antonelli}.

\begin{figure*}[]
\centering
\includegraphics[width=160mm]{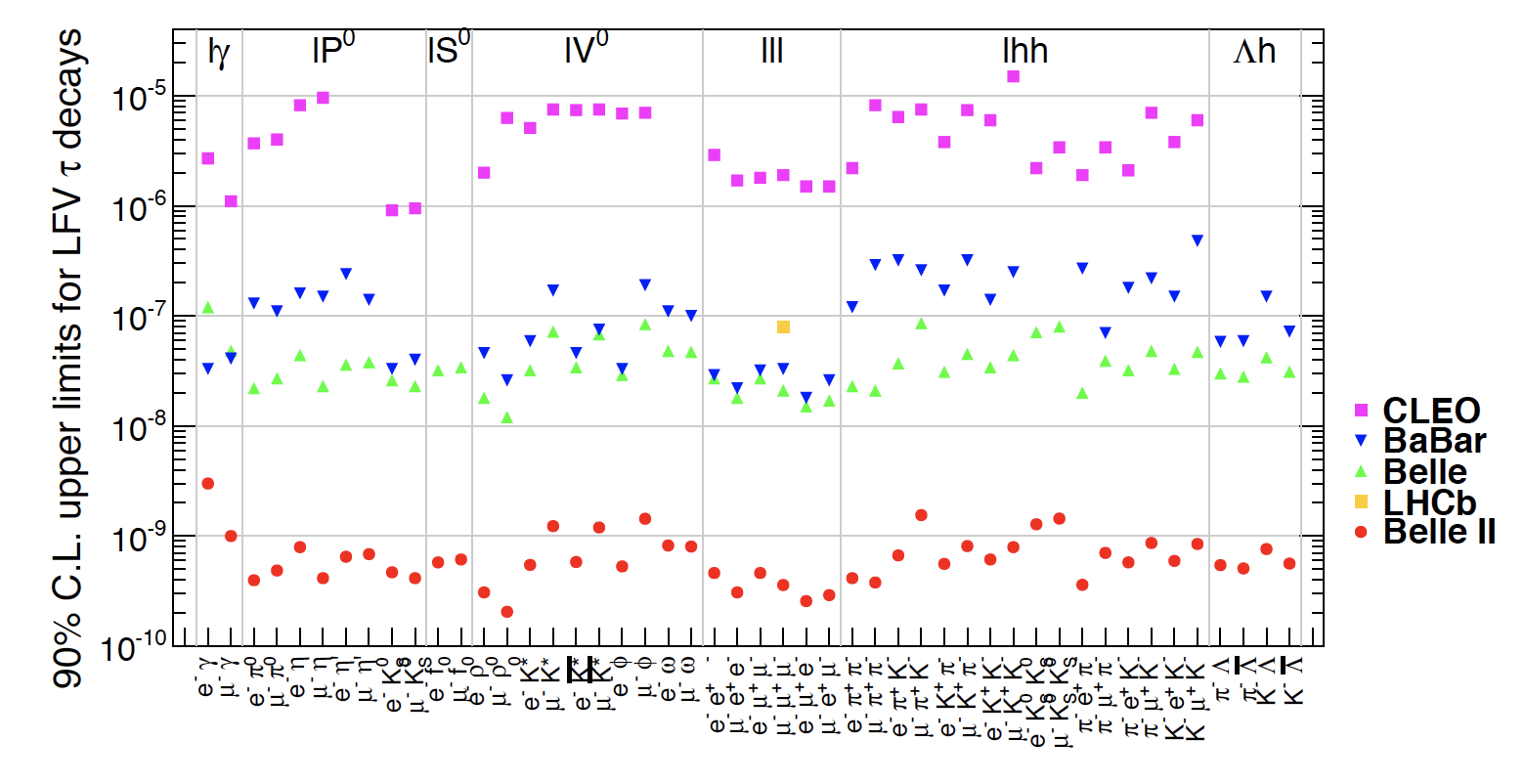}
\caption{ The branching fraction upper limits at $90\%$ CL of $\tau$ lepton-flavor-violating decays for CLEO (magenta solid squares), \babar\ (blue downward triangles), Belle (green upward triangles), LHCB (orange squares) and expected results for Belle II for a luminosity of $50~ab^{-1}$ (red squares).
 }
\label{fig:lfvt}
\end{figure*}

\section{Conclusion and Outlook}

The Belle experiment observed the decay $\tau^- \ra \pi^- \nu_{\rm \tau} e^+ e^-$ with a $5.9\sigma$ excess measuring a branching fraction of ${\cal B}(\tau^- \ra \pi^- \nu_{\rm \tau} e^+ e^-)= (2.11\pm 0.19\pm 0.30)\times 10^{-5}$. In the $\tau^- \ra \pi^- \nu_{\rm \tau} \mu^+ \mu^-$ channel the significance of the excess is $2.8\sigma$. So they set a branching fraction upper limit at $90\%$ CL of ${\cal B}(\tau \ra \nu_{\rm \tau} \mu^+ \mu^-) <1.14 \times 10^{-5}$. 
\babar\ measured the branching fractions of six signal channels $ \tau^-\ra K^- (0-3) \pi^0 \nu_{\rm \tau}$ and $\tau^- \ra-\pi^-( 3,4) \pi^0 \nu_{\rm \tau}$.
The new \babar\ results are the most precise except for $\tau^- \ra K^- \nu_{\rm \tau}$, which was previously measured with a three-prong tag. They help reducing the uncertainty on $|V_{\rm us}|$ determined from inclusive $\tau^- \ra X_{\rm s}^- \nu_{\rm \tau}$ decays. The new 
$|V_{\rm us}|$ result shows a $-2.9\sigma$ discrepancy  with respect to $|V_{\rm us}|$ determined from CKM unitarity.  
The $|V_{\rm us}|$ value extracted from the previous \babar\ $\tau^- \ra K^- \nu_{\rm \tau}$ measurement~\cite{babar10} is consistent with 
 the results from the CKM unitarity to better than $2\sigma$.
\babar\  measured the $\tau^- \ra K^- K^0_{\rm S} \nu_{\rm \tau}$  branching fraction, which is in excellent agreement with  the Belle measurement. The extracted spectral function is much more precise than  the measurement by CLEO. 

\babar\  will publish the $\tau^- \ra K^- n \pi^0 \nu_{\rm \tau}$ results, measure spectral functions in other $\tau^-$ decay modes and improve branching fraction measurements for other modes that are  relevant for improving the precision on $ |V_{us}|$ from $\tau^- \ra X_{\rm s}^- \nu_{\rm \tau}$ decays. The BES III experiment is working on a new $\tau$ mass measurement using 5 energy points at the $\tau$  threshold with a total integrated luminosity of $173~\rm  pb^{-1}$ expecting a mass precision of $\sigma (m_{\rm \tau}) < 100 ~\rm keV/c^2$.
The Belle II experiment will log a luminosity of $50~\rm  ab^{-1}$ yielding $4.6\times 10^{10}~ \tau$ pairs that allow for many improved $\tau$ measurements and many rare $\tau$ decay searches. Figure~\ref{fig:lfvt} shows the expected branching fraction upper limits at $90\%$ CL for various lepton-flavor-violating $\tau$ decays. The expected Belle II results~\cite{belle2}  will be two orders of magnitude or more lower than the present \babar\ and Belle results.




\begin{acknowledgments}
I would like to thank the \babar\ Collaboration for the opportunity to give this talk and Marcello Piccolo, Banerjee Swagato, Alessandre Filippi and Ian M Nugent Was for reviewing the slides. I would like to thank Alberto Luisiani and Tom Browder for supplying material  as well as Frank Porter and Shohei Nishida for checking the proceedings. 
\end{acknowledgments}


\begin{thebibliography}{9}
\bibitem{ccmode} Charge-conjugated modes are implied throughout this article.
\end{thebibliography}
\end{document}